\documentclass[aps,pre,twocolumn,showpacs,10pt]{revtex4-1}

\usepackage{graphicx}
\usepackage{dcolumn}
\usepackage{bm}
\usepackage{amsmath}
\usepackage{amssymb}
\usepackage{float}
\usepackage[titletoc,title]{appendix}

\begin{document}

\title{Topological pumping of photons in nonlinear resonator arrays}

\author{Jirawat Tangpanitanon\textsuperscript{1}}
\email{a0122902@u.nus.edu}
\author{Victor M. Bastidas\textsuperscript{1}}

\author{Sarah Al-Assam\textsuperscript{2}}

\author{Pedram Roushan \textsuperscript{3}}

\author{Dieter Jaksch\textsuperscript{1,2,4}}

\author{Dimitris G. Angelakis\textsuperscript{1,4}}
\email{dimitris.angelakis@qubit.org}

\affiliation{\textsuperscript{1}%
	Centre for Quantum Technologies, National University of Singapore, 3 Science Drive 2, Singapore 117543%
}
\affiliation{\textsuperscript{2}%
	Clarendon Laboratory, University of Oxford, Parks Road, Oxford OX1 3PU, United Kingdom%
}

\affiliation{\textsuperscript{3}%
Google Inc., Santa Barbara, California 93117, USA
}

\affiliation{\textsuperscript{4}%
	School of Electronic and Computer Engineering, Technical University of Crete, Chania, Crete, Greece, 73100%
}

\date{\today}

\begin{abstract}

We show how to implement topological or Thouless pumping of interacting photons in one dimensional nonlinear resonator arrays, by simply modulating the frequency of the resonators periodically in space and time.  The interplay between interactions and the adiabatic modulations enables robust transport of Fock states with few photons per site. We analyze the transport mechanism via an effective analytic model and study its topological properties and its protection to noise.  We conclude by a detailed study of an implementation with existing circuit QED architectures.

 \end{abstract}
 
 
\maketitle


\textit{Introduction.---} In the 3rd century BC, Archimedes figured out how to pump water up a hill using a rotating screw simply by exploiting the laws of classical physics. Two millenia later,  Thouless  proposed topological pumping for the robust transport of quantum particles in 1D periodic lattices, by means of an analogous adiabatic and cyclic deformation of the underlying Hamiltonian~\cite{thouless1983}.  In addition, he showed that the number of particles transported during one pump cycle is quantized and can be related to a topological invariant, known as the Chern number~\cite{topologicalInsulator}. As a consequence, the transport is robust against small disorder and perturbation~\cite{Thouless1984,Seiler1985,Kwek2014,PhysRevB.91.125411,Gong2015,Gong2016}. 

Recently, there have been experimental demonstrations of such topological or ``Thouless" pumping for non-interacting particles using  optical lattices~\cite{bloch2016, takahashi2016} and waveguide arrays~\cite{kraus2012,Silberberg2015}.
Topological pumping in the case of interacting systems remains relatively unexplored. Previous works have theoretically explored related adiabatic quantum pumping in quantum wires~\cite{Niu2003}, quantum dots~\cite{Andreev1998,Fazio2005,Silva2008}, and electrons in a mesoscopic conductor~\cite{Oreg2006}.  
However, the latter schemes do not involve a lattice structure which is essential for achieving topological protection. In the case of interacting systems in a 1D lattice, topological pumping has been explored formally in the bosonic Bose-Hubbard model with correlated hopping and nearest-neighbor repulsion~\cite{Altman2011,Fazio2013}, and in interacting spin systems~\cite{Brouwer2011}.

In this work, we propose a realistic implementation of topological pumping with interacting photons in nonlinear resonator arrays ~\cite{dimitrisreview, dimitris2007, hartmann2006,greentree2006}. Complementary to the previous experiments with non-interacting particles~\cite{kraus2012,Silberberg2015,bloch2016, takahashi2016}, in our model, the interplay between interactions and the adiabatic modulations enables robust transport of Fock states, with few photons per site. In other words, our pumping scheme allows the transport of several-body  particle (photon) states in the strongly correlated regime and thus it is qualitatively different from the standard Thouless pumping approaches ~\cite{Thouless1984,Seiler1985,Kwek2014,PhysRevB.91.125411,Gong2015,Gong2016}. Robustness of the transport against disorder is analyzed numerically for a range of parameters. Beyond numerics, we also discuss an effective model, which provides a simple intuition of the underlying mechanism of our transport scheme  in the nonlinear regime. Finally, by  solving the corresponding Lindblad master equation for a Circuit QED array of 9 sites~\cite{Houck, Devoret,PhysRevLett.113.220502, google}, we analyze how the quantized transport of a three-photon Fock state could be observed in such existing experimental set-ups.

\begin{figure}
\centering
\includegraphics[width=8.5cm,height=8.3cm]{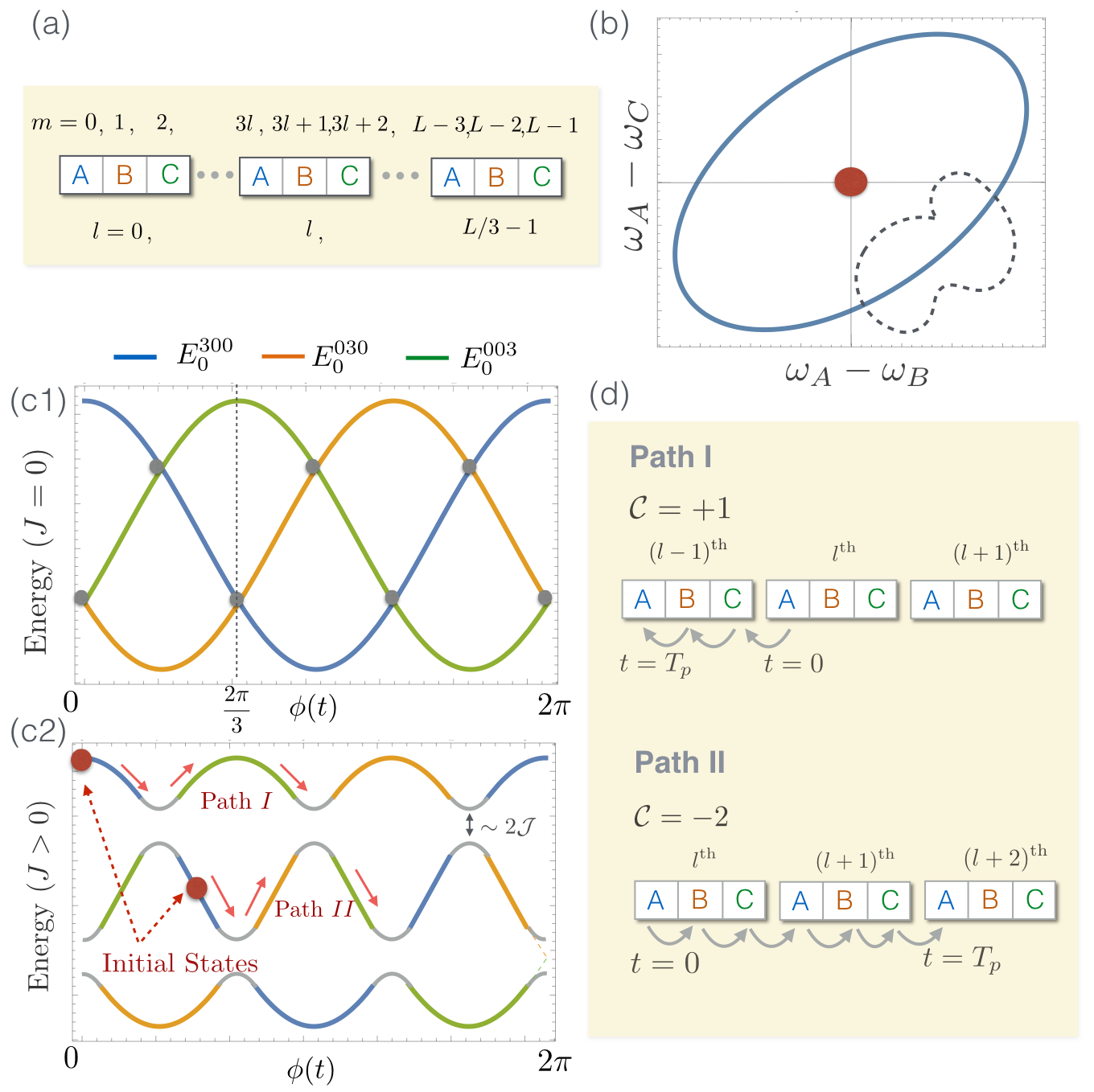}
\caption{(a) Depicts of the sublattices $A$, $B$, and $C$ at the sites $3l$, $3l+1$ and $3l+2$, respectively. (b) Pump cycle in the 2D parameter space spanned by ($\omega_A-\omega_B$) and ($\omega_A-\omega_C$) for $U=-J$. It encircles the critical point at $\omega_A=\omega_B=\omega_C$, labelled as a red dot. A gray-dashed path is displayed as an example of a topologically-trivial path. (c1) On-site energies, $E^{\mu}_{0}(t)=\langle \mu |H^l_0(t)|\mu\rangle_l$, as a function of the modulation phase $\phi(t)$. Different bands $\mu=300,030,003$ are labelled as blue, green and orange, respectively. Crossing points between two bands are labelled as grey dots. (c2) Eigenenergies emerging in the presence of a small photon hopping $J\ll \Delta$. As discussed in the text, near every crossing points in (c1) an effective three-photon hopping can be derived, which convert these points into the anti-crossing points, shown in (c2) with the gap $2 \mathcal{J}=\sqrt{2}J^3/U^2$.  As a result, the quantized transport of the Fock states can then be understood by adiabatically following one of the bands in (c2). (d) Illustration of the quantized transport. $T_p=2\pi/\Omega$ is the pumping period. In the path $I$, the state $| 300\rangle_l$ is initialized at the highest band with $\phi(0)=0$. The three photons hop from one site to another when passing through each anti-crossing point. Since in the upper band there are three anti-crossing points for $\forall\phi(t)\in\left[0,2\pi\right)$. After one pump cycle, the three photons are pumped from  $| 300\rangle_l$ to  $|300\rangle_{l-1}$. The transport corresponds to the effective Chern number $\mathcal{C}=1$. For the path $II$, the transport has different topology with $\mathcal{C}=-2$. The lowest band has the same topology as the highest one.}
\label{fig:fig1}
\end{figure}

\textit{The system. ---} We consider a nonlinear resonator array of size $L$, implemented in Circuit QED as discussed in the Supplemental Material~\cite{SupplementalInfo}. The array is described by the Bose-Hubbard model (BHM) with attractive interactions and spatially modulated on-site energies \footnote{In the semi-classical regime \unexpanded{$\langle\hat{n}_m\rangle\gg1$}, the Hamiltonian in Eq.~\eqref{eq:harper} is known as the nonlinear Harper model. Previous works have used the mean-field description to study the spreading of an initially-localized wave packet~\cite{ng2007,lahini2009,manela2010,zhang2011,larcher2012}. In the absence of interactions \unexpanded{$U=0$}, the model can be mapped to the 2D integer quantum Hall system. The latter has been realized in optical waveguide arrays ~\cite{kraus2012}.}
\begin{align}
      \label{eq:harper}
            H(t) = &\sum_{m=0}^{L-1}\omega_m(t) \hat{n}_m -J \sum_{m=0}^{L-2} \left(\hat{a}^\dagger_m\hat{a}_{m+1}+\text{H.c.}\right)\nonumber \\ &+\frac{U}{2}\sum_{m=0}^{L-1}\hat{n}_m(\hat{n}_m-1),
\end{align}
where $\hat{n}_m=\hat{a}_m^{\dagger}\hat{a}_m$ and $\hat{a}^{\dagger}_m$, $\hat{a}_m$ are bosonic creation and annihilation operators, respectively. In addition, 
$\omega_m(t)=\omega_0+\Delta\cos\left[2\pi m/3 +\phi(t) \right]$ is the frequency of the resonator, $\Delta>0$ is a modulation amplitude,  $\phi(t)$ is a time-dependent modulation phase, $U<0$ is an attractive Kerr nonlinearity,  and $J>0$ is the hopping strength. Since the number of photons is conserved, the first term $\sum^{L-1}_{m=1}\omega_0\hat{n}_m$ can be eliminated from Eq~\eqref{eq:harper}. The modulation phase is adiabatically swept in time as $\phi(t)=\Omega t +\phi_0$, where $\Omega$ is the ramping speed and $\phi_0$ is an initial modulation phase. This sweeping is possible in circuit QED, as the frequency of the resonator can be tuned on the fly using a flux bias ~\cite{PhysRevLett.113.220502,transmon}.
We provide more details on the implementation in the last section and also in the Supplemental Material~\cite{SupplementalInfo}. 
For simplicity, we choose $L$ to be an integer multiple of three. Hence, the array can be thought of as an array of trimers of size $L/3$ as it is depicted in Fig.~\ref{fig:fig1}(a). 
For convenience, we introduce the index $l\in\{0,\dots,L/3-1\}$ to label the trimers. Since the modulation wavelength is also three, the Hamiltonian that acts on each trimer is identical. Later in the text, we will refer to the three sublattices at positions $3l$, $3l+1$ and $3l+2$ within the $l$-th trimer as $A$, $B$, and $C$, respectively [see Fig.~\ref{fig:fig1}(a)].

Our pumping protocol is shown in Fig.~\ref{fig:fig1}(b). It corresponds to a loop in a 2D parameter space with $U=-J$. Our two varying parameters are the frequency differences $\omega_A-\omega_B$ and $\omega_A-\omega_C$ between two resonators in a trimer, where $\omega_A=\omega_0+\Delta\cos(\phi(t))$, $\omega_B=\omega_0+\Delta\cos(\phi(t)+2\pi/3)$, and $\omega_C=\omega_0+\Delta\cos(\phi(t)+4\pi/3)$. In the Supplemental Material~\cite{SupplementalInfo}, we show that for $U=-J$ this loop encloses the critical point when $\omega_A=\omega_B=\omega_C$. We will later show that the spectrum along this loop is gapped. As a result, the topology of the pump is said to be invariant under deformation of this loop, as long as it encloses the critical point~\cite{topologicalInsulator}.


\textit{Topological transport of Fock states.---} Let us begin by considering three-photon Fock states within a given trimer, i.e., $|300\rangle_l$, $|030\rangle_l$ and $|003\rangle_l$. In the following, we will show that at the right regime an effective three-photon hopping can be derived, allowing the three-photon Fock states to be efficiently transported through the array.   
  
To illustrate the above, let us decompose the Hamiltonian as $H(t) = \sum_{l} H_0^l(t) + \lambda V$, where $H_0^l(t)=\sum_{m=3l}^{3l+2}\left[\Delta \cos[2\pi m/3+\phi(t)] \hat{n}_m +\frac{U}{2}\hat{n}_m(\hat{n}_m-1)\right]$ and $\lambda V= - J \sum_m \left(\hat{a}^\dagger_m\hat{a}_{m+1}+\text{H.c.}\right)\nonumber$. In the case $J=0$, we define the on-site energies of the three-photon Fock states as $E^{\mu}_{0}(t)=\langle \mu |H^l_0(t)|\mu\rangle_l$ for $\mu\in\{300,030,003\}$. Note that the energies $E^{\mu}_{0}(t)$ do not depend on the trimer index $l$. The energies $E^\mu_0(t)$ are depicted in Fig.~\ref{fig:fig1}(c1) as a function of $\phi(t)$. When including a small but nonvanishing hopping strength $J\ll \Delta$, the crossings in Fig.~\ref{fig:fig1}(c1) become anticrossings, as shown in Fig.~\ref{fig:fig1}(c2). This is due to an effective three-photon hopping, that happens near an anticrossing (we outline the mechanism below). As a result, the energy spectrum when $0<J\ll \Delta$ is separated into three gapped bands for $ \forall\phi(t)\in \left[0,2\pi\right)$, as depicted in Fig.~\ref{fig:fig1}(c2). The quantized transport of the three photons can then be understood simply by adiabatically following one of these bands, [see Figs.~\ref{fig:fig1}(c2) and (d)].

To obtain the above picture, we derive the effective three-photon hopping by first identifying relevant states near a given anticrossing point. To do this let us consider a particular crossing point in Fig.~\ref{fig:fig1}(c1) when $J=0$, for example at $\phi(t^*)=2\pi/3$. There, the two crossing bands, $E^{300}_0(t^*)=E^{030}_0(t^*)$, are far-separated from the third one, $E^{003}_0(t^*)$. Hence when including a small hopping $J\ll \Delta$, the relevant three photon states in the $l$-th trimer are $|300\rangle_l$, $|030\rangle_l$, $|210\rangle_l$ and $|120\rangle_l$. We then note that when $J=0$ the states $|300\rangle_l$ and $|030\rangle_l$ have the same on-site energies, $\epsilon_3=E^{300}_0(t^*)=E^{030}_0(t^*)=-3\Delta/2+3U$, and so do the states $|210\rangle_l$ and $|120\rangle_l$ with the on-site energy $\epsilon_2=-3\Delta/2+U$. Since $\epsilon_2-\epsilon_3=-2U$, in the limit $0<\sqrt{3}J<-2U$, one can do adiabatic elimination of the intermediate states $|210\rangle_l$ and $|120\rangle_l$, during the process $| 300\rangle_l\to|210\rangle_l\to|120\rangle_l\to|030\rangle_l$. This is done by expanding the Hamiltonian $H(t^*)$ in $\sqrt{3}J/2U$ up to the third order using the Schreffer-Wolff transformation~\cite{cohen1998}. As a result, an effective three-photon hopping process can be derived  as $\tilde{H}^l_{J}(t^*)=-\mathcal{J}(|300\rangle_l\langle 030|_l+|030\rangle_l\langle 300|_l)$, where $\mathcal{J}=J^3/\sqrt{2}U^2$ (See Supplemental Material~\cite{SupplementalInfo}). Similar analysis can be applied for all anti-crossing points in Fig.~\ref{fig:fig1}(c2). We stress here that this perturbation does not work in the absence of interactions, i.e., when $U=0$.



The three gapped bands in Fig.~\ref{fig:fig1}(c2), resulting from the effective three-photon hopping, are said to have different topologies due to their distinct transport properties. For example, as shown in Fig.~\ref{fig:fig1}(d), the states in the middle band move in the opposite direction with twice the speed as those in the upper band. We define the effective Chern number $\mathcal{C}$---a topologically invariant quantity for a given band---as the number of trimers that the three photons pass during one pump cycle, which is equivalent to the Wannier center displacement~\cite{topologicalInsulator}. Hence, the states in the highest and the middle bands can be assigned with the Chern numbers $\mathcal{C}=+1$ and $\mathcal{C}=-2$, respectively. The sign indicates whether the direction of motion is the same $(+)$ as or opposite $(-)$ to that of the pump. (Recall that the modulation wave $\Delta\cos(2\pi m/3+\Omega t+\phi_0)$ moves towards the leftmost site $m=0$). 



\begin{figure}
\centering
\includegraphics[width=8.7cm,height=3.7cm]{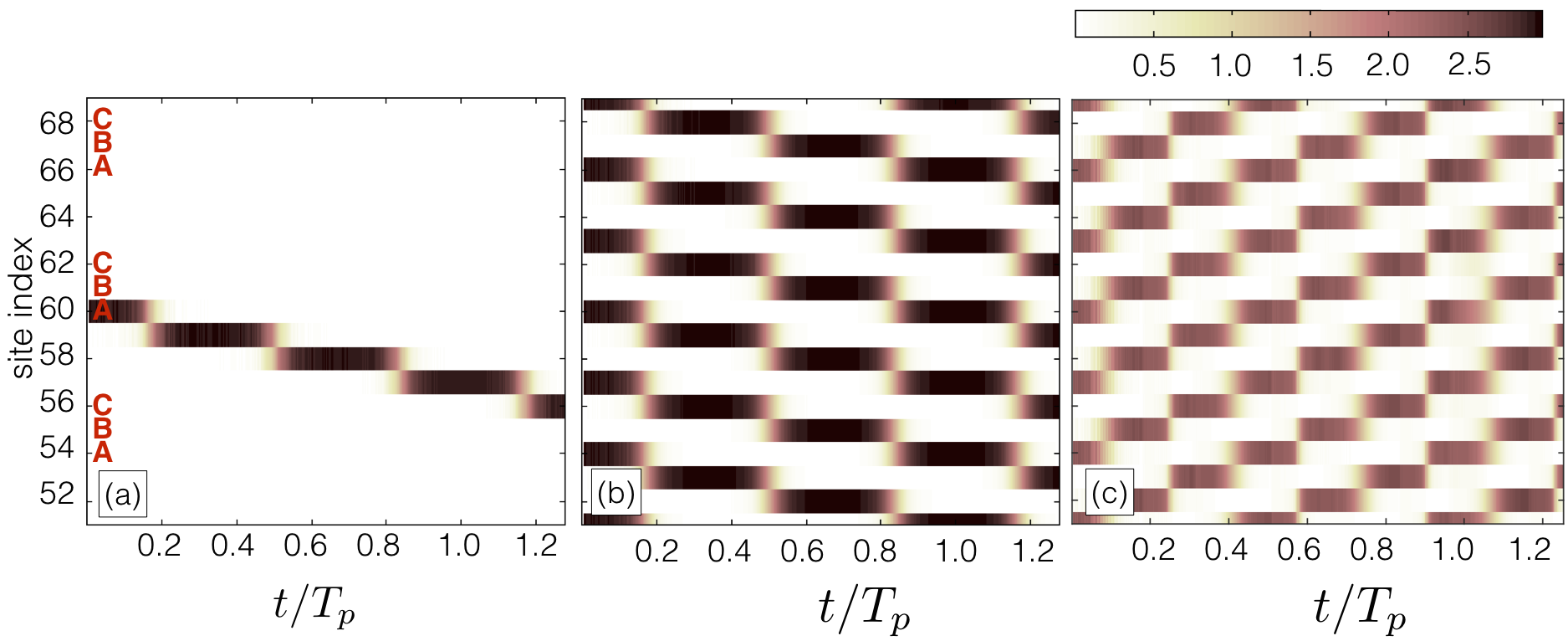}
\caption{Density plot $\langle \hat{n}_m\rangle$ as a function of time, illustrating a quantized transport of a three-photon state. In (a), a three-photon Fock state $|3\rangle$ is prepared at the sublattice A, located at the middle of an array of size $L=120$ ($\Delta=10J$, $\phi_0=0$, $U=-J$, $T_p=2\pi/\Omega$, and $\Omega=0.01J$). In (b), each sublattice A is filled with the three-photon Fock state. We left five trimers near the edges empty to avoid boundary effects during the evolution. The density plot shows a clear step-like behavior in both cases. In (c), the initial modulation phase is set at $\phi_0=\pi/2$ and the ramping speed is $\Omega_p=0.002J$. As discussed in the text, this results in a quantized transport in the reversed direction and twice the speed of the pump. In (b) and (c), the local Hilbert space in the numerics is truncated at the five photon Fock state .}
\label{fig:fig2}
\end{figure}

In Fig.~\ref{fig:fig2}(a), we numerically show the quantized transport by plotting the density $\langle \hat{n}_m\rangle$ as a function of time. The three-photon Fock state is initialized at the site $m=60$ (sublattice $A$) of an array of size $L=120$. Time evolution is performed using Time-evolving Block Decimation (TEBD) ~\cite{tnt, PhysRevLett.91.147902} with bond dimension 100. The parameters of the Hamiltonian are $\Delta=10J$, $U=-J$, $\Omega=0.01J$ and $\phi_0=0$. The density plot shows a clear step-like motion with $\mathcal{C}=1$, as expected. 

One can also imagine filling each trimer $l$ with the same three-photon Fock states $| 300\rangle_l$, i.e. the unit-filling condition [see Fig.~\ref{fig:fig2}(b)]. Because photons between neighboring trimers are always two-sites apart, they are effectively decoupled from each other throughout the evolution. Hence, a similar quantized transport occurs for many-photon states.

\begin{figure}
\centering
\includegraphics[width=8.5cm,height=4.8cm]{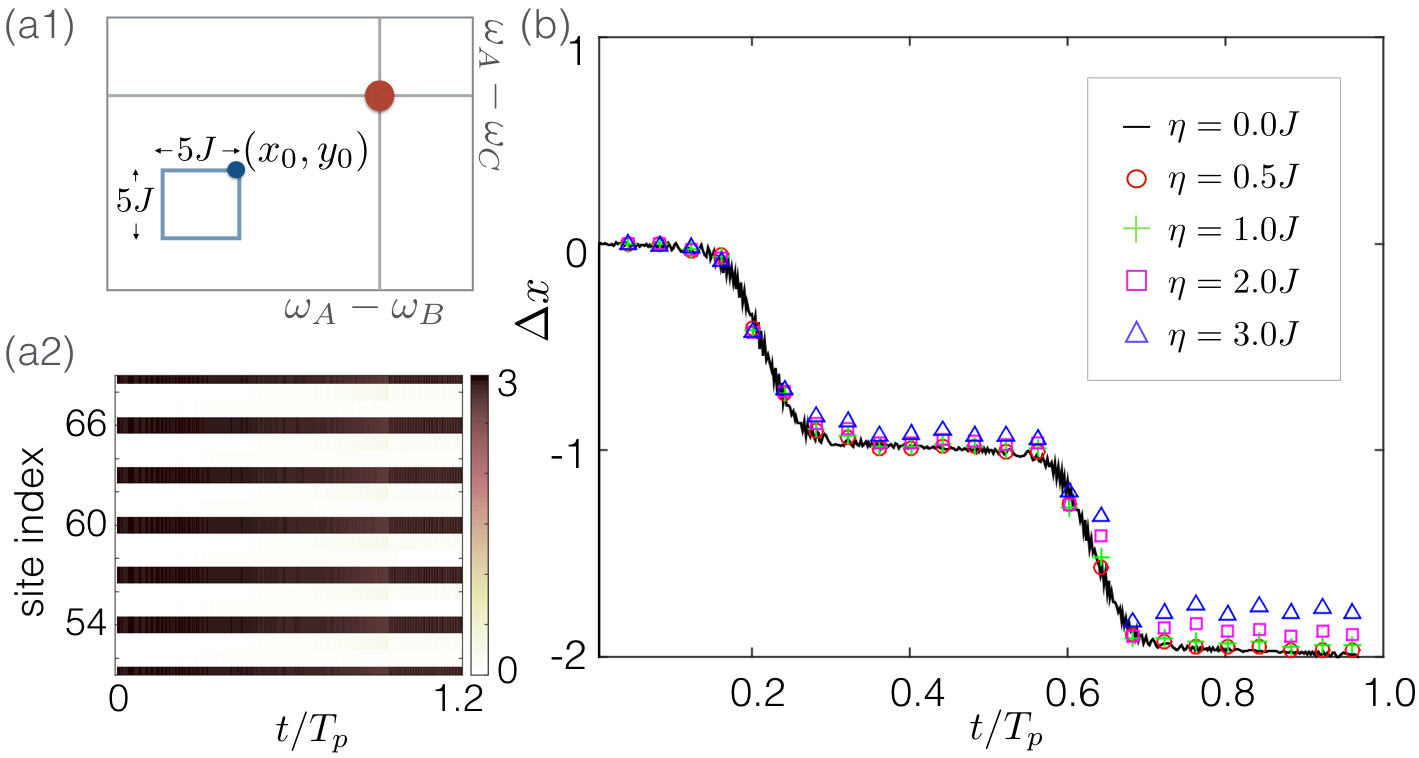}
\caption{(a1) Illustration of a topologically-trivial pumping scheme in the 2D parameter space. We fix $\omega_A=\omega_0$ and vary $\omega_B$ and $\omega_C$ as a square loop with the starting point $(x_0,y_0)=(-\Delta \cos (2\pi/3),-\Delta \cos (4\pi/3))$. The pumping period $T_p$ and the initial state are the same as those in Fig.\ref{fig:fig2}(b). (a2) Density plot showing the corresponding motion. (b) Center-of-mass (COM) displacement, $\Delta x$, of a three-photon state as a function of time with the non-trivial pumping topology in the presence of random noise. A black solid line corresponds to the perfect case with no noise $\eta=0$. The parameters of the Hamiltonian and the initial state are the same as those in Fig.\ref{fig:fig2}(a). The plot shows that the quantized motion is robust against weak perturbations, such that the amplitude of the noise, $\eta$, is smaller than the smallest energy gap $2\mathcal{J}$.}
\label{fig:fig4}
\end{figure}

The quantized transport with $\mathcal{C}=-2$ is shown in Fig.~\ref{fig:fig2}(c), where the initial modulation phase is changed to $\phi_0=\pi/2$ while keeping the initial state the same as that in Fig.~\ref{fig:fig2}(b). The motion is reversed with twice the speed compared to the one in Fig.~\ref{fig:fig2}(b), as expected. To further emphasize the topological nature of the transport, in Fig.~\ref{fig:fig4}(a1) we also change our pumping scheme to the one that does not enclose the critical point in the 2D parameter space, while keeping the starting and the end points the same as before. As shown in Fig.~\ref{fig:fig4}(a2), photons following this path remain localized throughout the evolution, corresponding to a topologically-trivial transport with $\mathcal{C}=0$, as expected.


\textit{Robustness analysis .---} The quantized transport discussed here so far is protected by the energy gap proportional to $2 \mathcal{J}$. Hence, it is robust against small perturbations. To illustrate this, we add random noise to the system as $H_{\text{noise}}=\eta \sum_mr_{m}(t) \hat{n}_m$, where $\eta$ is the noise amplitude and $r_{m}(t)\in \left[0,1\right]$ is a random number, drawn differently for each site at each time step. The parameters of the Hamiltonian and the initial state are the same as in Fig.~\ref{fig:fig2}(a). The center-of-mass (COM) of the three photons as a function of time, with an increasing $\eta$, is shown in Fig.~\ref{fig:fig4}. It can be seen that the quantized motion is robust against weak perturbations, $\eta \lesssim J$. As the noise amplitude $\eta$ becomes comparable to the smallest energy gap, which in this case is $2\mathcal{J}\sim 1.4 J$, the COM is biased towards the rightmost site ($m=L-1$). This is expected, as random noise introduces coupling to states from other bands. As shown before, these states are transported in the opposite directions. Therefore when $\eta\gtrsim\mathcal{J}$, the COM deviates from the ideal case over time.

\begin{figure}
\centering
\includegraphics[width=8.8cm,height=6.8cm]{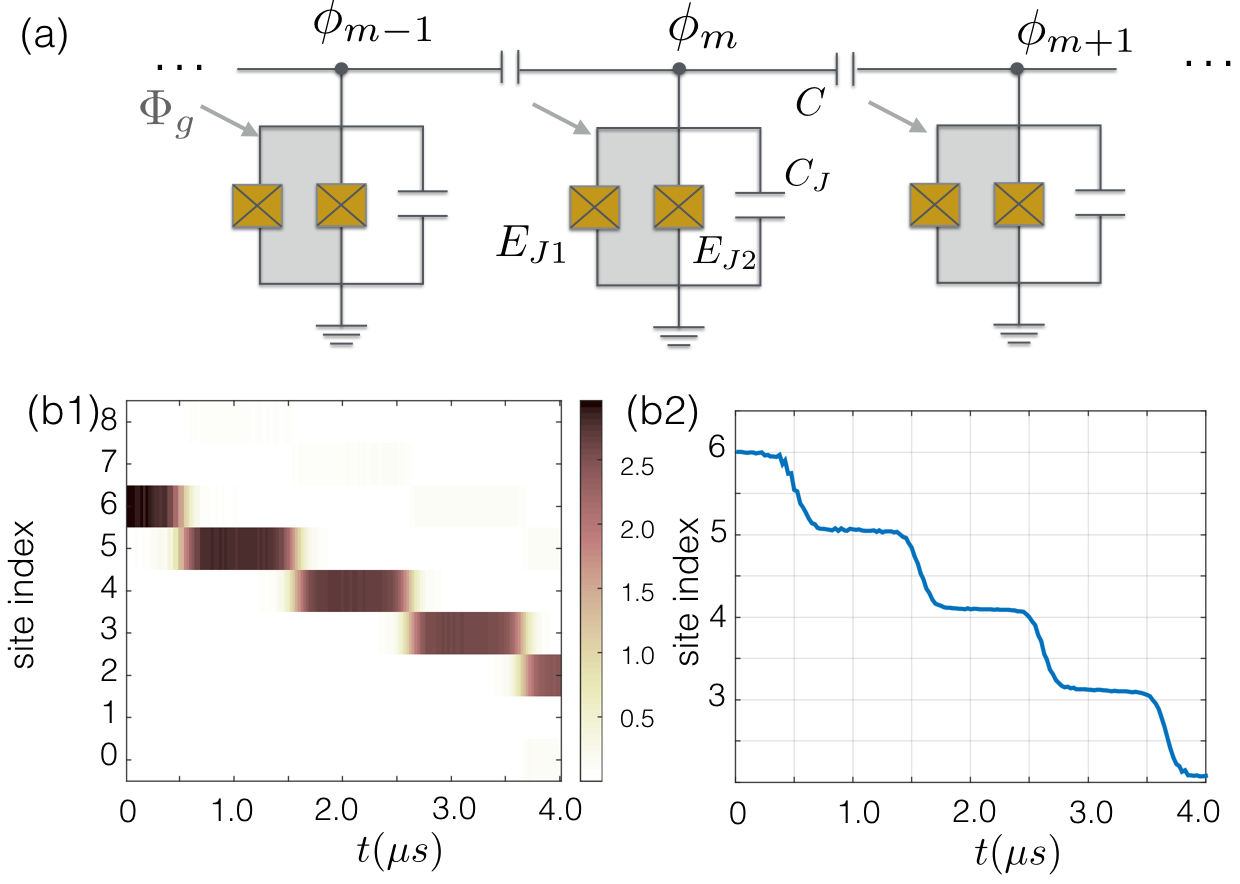}
\caption{(a) Circuit QED diagram showing an implementation of the Hamiltonian $H(t)$. We introduce the flux variable which is defined as $\phi_m=-\int V_m dt$, where $V_m$ is a voltage at the corresponding position. As shown in the Supplemental Material, this quantity can be quantized to the form $\phi_m=\alpha (a_m+a^\dagger_m)$, where $\alpha$ is a constant depending on the circuit's elements. The Josephson junctions $E_{J1}, E_{J2}$ and the shunting capacitor $C_J$ acts as a nonlinear resonator, whose frequency can be tuned via the flux bias $\Phi_g$. Each resonator is coupled by the capacitor $C$. (b1)-(b2) Quantum trajectory simulations of a nine-site lossy resonator array. (b1) Density plot $\langle\hat{n}_m\rangle$ as a function of time in a lossy case. (b2) Center-of-mass of the three photons as a function of time. A clear step-like behavior is observed in both plots. }
\label{fig:fig5}
\end{figure}


\textit{Circuit QED implementation.---} The localization due to attractive interaction and large modulation $\Delta\gg J$ in our system allow signatures of topological pumping to be visible with an existing finite-size array, as small as $L=9$~\cite{google}. (See Fig.\ref{fig:fig5}(a) and also Supplemental Material~\cite{SupplementalInfo} for more details on the implementation). To show this, we numerically solve the Lindblad master equation involving realistic photon loss, which is expressed as
\begin{equation}
\frac{\partial \rho}{\partial t}=-i[H(t),\rho]+\frac{1}{2T_1}\sum_{m}\left(2\hat{a}_m\rho \hat{a}^\dagger_m-\{\hat{n}_m,\rho\}\right),
\end{equation}
where $\rho$ is a density matrix and $T_1=20$ $\mu s$ is a photon lifetime. We choose realistic parameters of the Hamiltonian as $\Delta=0.4$ GHz, $\Omega=2$ MHz, $J=40$ MHz and $U=-40$ MHz. Three photons are initialized at the site $m=6$ with $\phi_0=0$. Time evolution is performed by averaging over 1000 quantum trajectories \cite{daley}. The density and the COM plots as a function of time are shown in Fig.~\ref{fig:fig5}(b1) and Fig.~\ref{fig:fig5}(b), respectively. A clear step-like motion is observed in both plots.


In conclusion, we have proposed a realistic realization of topological pumping of photons in the interacting regime using coupled resonator arrays. The available interactions enable the topological transport of few photon Fock states robustly against disorder. We also discuss in detail how to  implement the above in existing circuit QED architectures. 

We note that although we have been focusing on the transport of Fock states,  entangled states are also created and can be transported through by adjusting the initial conditions. Hence, in the future it would be interesting to charecterize this entanglement, which emerges between the two neighbouring sites during population transfer, and seek applications in quantum information processing.

\textit{Acknowledgements.---} The authors acknowledge fruitful discussions with P. Zoller. The authors gratefully acknowledge financial support through the National Research Foundation and Ministry of Education Singapore (partly through the Tier 3 Grant ``Random  numbers  from  quantum  processes''); and travel support by the EU IP-SIQS.%


\begin{widetext}

\section*{Supplemental Material for ``Topological pumping of photons in nonlinear resonator arrays''}

Equations in the main paper are denoted by Eq.~[*].

\subsection{Effective three-photon hopping.\label{AppendixA}}

In this section, we review the Schrieffer-Wolff (SW) transformation \cite{cohen1998} and use it to derive an effective three-photon hopping process. As discussed in the main text, we restrict ourselves to a three-photon manifold of the $l$-th trimer, i.e. $|300\rangle_l$, $|030\rangle_l$, $|210\rangle_l$ and $|120\rangle_l$. In this subspace, the Hamiltonian is decomposed as 
\begin{equation}
H_l=H^{0}_{l}+\lambda V_l, 
\end{equation}
where $H^{0}_l = \sum_{m=3l}^{3l-2}\left[\Delta \cos (2\pi m/3+\phi(t^*))+n_{m}(n_{m}-1)\right]$ and $\lambda V_l = -J\sum_{m=3l}^{3l-2}\left(a^\dagger_{m}a_{m+1}+\text{H.c.}\right)$, with $t^*=2\pi/3$.

When $J=0$, the states $|300\rangle_l$ and $|030\rangle_l$ have the same on-site energy $\epsilon_3$ and so do the states $|210\rangle_l$ and $|120\rangle_l$ with the energy $\epsilon_2$. Hence, the spectrum of $H^0_{l}$ can be grouped into two manifolds, labelled as $\mathcal{D}_3=\{|300\rangle_l,|030\rangle_l\}$ and $\mathcal{D}_2=\{|210\rangle_l,|120\rangle_l\}$ respectively (see Fig.\ref{fig:a1}). The two manifolds are separated by a gap $\Delta E = \epsilon_2-\epsilon_3=-2U$. We consider the hopping term $\lambda V_l$ as a perturbation that couples these manifolds. 

\begin{figure}
\includegraphics[width=5cm,height=3.2cm]{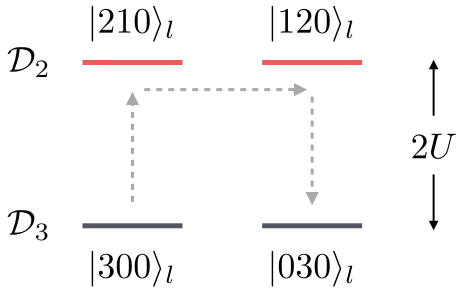}
\caption{Diagram showing a third-order three-photon hopping process. For the $l$-th trimer, the states $|300\rangle_l$ and $|030\rangle_l$ have the same on-site energy as well as $|210\rangle_l$ and $|120\rangle_l$. Hence, they can be grouped into two manifolds, labeled as $\mathcal{D}_3$ and $\mathcal{D}_2$ respectively. The two manifolds are separated by $\Delta E = 2U$. Since the hopping element between $|210\rangle_l$ and $|300\rangle_l$ is $\sqrt{3}J$, we have $\sqrt{3}J/\Delta E< 1$ for $U=J$. This allows a relatively strong third-order hopping process where the three photons hop from $|300\rangle_l$ to $|030\rangle_l$ via the intermediate states $|210\rangle_l$ and $|120\rangle_l$ and vice versa. }
\label{fig:a1}
\end{figure}

Our aim is to find an effective Hamiltonian $H'_l$ which has no matrix elements between the two manifolds. We require that the effective Hamiltonian is related to the original Hamiltonian by a unitary transformation $H'_l=e^{\mathrm{i}S^l}He^{-\mathrm{i}S^l}$, where $S$ is a Hermitian matrix. By expanding $S^l=\lambda S^l_1+\lambda^2S^l_2+\lambda^3S^l_3+...$, the effective Hamiltonian can be expressed up to the third order in $\lambda$ as $H'_l=H_l^0+H_l'^{(1)}+H_l'^{(2)}+H_l'^{(3)}$, where
\begin{align}
H'^{(1)}_l=&[\mathrm{i}\lambda S^l_1,H_l^0] + \lambda V_l,  \nonumber \\
H'^{(2)}_l=&[\mathrm{i}\lambda^2 S^l_2,H_l^0]+[\mathrm{i}\lambda S^l_1, \lambda V_l] + \frac{1}{2}[\mathrm{i}\lambda S^l_1,[\mathrm{i}\lambda S^l_1,H^0_l]], \nonumber \\
H'^{(3)}_l=&[\mathrm{i}\lambda^3 S^l_3,H^0_l]+[\mathrm{i}\lambda^2S^l_2,\lambda V_l] + \frac{1}{2}[\mathrm{i}\lambda^2S^l_2,[\mathrm{i}\lambda S^l_1,H_l^0]] + \frac{1}{2}[\mathrm{i}\lambda S^l_1,[\mathrm{i}\lambda^2(S^l)^2_2,H^0_l]]  \nonumber \\ 
              &+\frac{1}{2}[\mathrm{i}\lambda S^l_1,[\mathrm{i}\lambda S^l_1,\lambda V_l]] + \frac{1}{6}[\mathrm{i}\lambda S^l_1,[\mathrm{i}\lambda S^l_1,[\mathrm{i}\lambda S^l_1,H^0_l]]].
\end{align}

Let us consider the first order term $H'^{(1)}_l$. Since the matrix element of $H'_l$ between two manifolds is zero, we have $\langle \alpha |H'^{(1)}_l|\beta\rangle=0$, where $|\alpha\rangle$ and $|\beta\rangle$ are states from  manifolds $\mathcal{D}_{\alpha}$ and $\mathcal{D}_{\beta}$ with $\alpha\neq\beta$, respectively. For example, one has that $\langle 300 |_lH'^{(1)}_l|210\rangle_l=0$.
However, the $S^l$ matrix that satisfies this condition is not unique. To avoid this, we choose the $S^l$ matrix such that it does not have matrix elements inside each manifold, i.e. $P_\alpha S^lP_\alpha=0$, for $ \alpha \in \{2,3\}$ where $P_{\alpha}$ is a projector over the manifold $\mathcal{D}_\alpha$. Therefore, one has $P_2=|210\rangle_l\langle 210|_l+|120\rangle_l\langle120|_l$ and $P_3=|300\rangle_l\langle300|_l+|030\rangle_l\langle030|_l$.

With this, the first-order matrix $S^l_1$ can be written as 
\begin{equation}
\mathrm{i}\lambda S^l_1 = \sum_{\alpha,\beta}\frac{\langle \alpha |\lambda V_l|\beta\rangle}{\epsilon_{\beta}-\epsilon_{\alpha}}|\alpha\rangle \langle \beta |=\frac{\sqrt{3}J}{2U}\left(|300\rangle_l \langle 210|_l +  |030\rangle \langle 120 |_l - h.c.\right).
\end{equation}
Similarity, since $\langle \alpha |H'^{(2)}_l|\beta\rangle=0$, the second-order matrix $S^l_2$ is
\begin{equation}
\mathrm{i}\lambda^2 S^l_2 = \sum_{\alpha,\beta}\frac{\langle \alpha | [\mathrm{i}\lambda S^l_1,\lambda V_l]|\beta\rangle}{2(\epsilon_{\beta}-\epsilon_{\alpha})}|\alpha\rangle \langle \beta| = \frac{\sqrt{3}J^2}{4\sqrt{2}U^2}\left(|300\rangle_l \langle 120|_l + |030\rangle_l \langle 210|_l - h.c. \right).
\end{equation}
The third order commutator $[\mathrm{i}\lambda^3 S^l_3,H^0_l]$ is off-diagonal and, by definition, does not contribute to the term $H'^{(3)}_l$. By restricting to the $\mathcal{D}_3$ manifold, i.e. $P_3H'_lP_3$, the three-photon hopping can be derived from the third-order term as
\begin{equation}
H'^{(3)}_{l,J} = -\frac{J^3}{\sqrt{2}U^2}(|300\rangle_l\langle 030|_l + |030\rangle_l \langle300 |_l).
\end{equation}
Other terms in $P_3H'_lP_3$ result in a normalization factor of the on-site energies.

\subsection{Mean field description \label{AppendixB}}
To discuss the critical properties of the model Eq.~[1], we resort on the mean field analysis~~\cite{HammererMarq2014,BastidasChim2016}. With this aim, we consider a lattice with $L$ sites and periodic boundary conditions $a_{0}=a_{L-1}$.
We introduce a new set of displaced bosonic operators
\begin{equation}
	a_{m}=b_{m}+\alpha_{m}
	\label{eq:D}
	\ ,
\end{equation}
where $a_{m}$ are the original bosonic operators,  and $b_{m}$ describes the quantum fluctuation about the mean field $\alpha_{m}$.

Let us focus now on the particular case of a time independent phase $\phi(t)=\phi_0$ in Eq.~[1] and $b=1/3$. In this particular case, the one dimensional lattice is composed by $L/3$ trimers with on-site energies
\begin{align}
      \label{eq:onsiteEn}
            \omega_{A}&=\omega_{3l}=\omega_0+\Delta\cos\phi_0
            \nonumber\\
            \omega_{B}&=\omega_{3l+1}=\omega_0+\Delta\cos(\phi_0+2\pi/3)
            \nonumber\\
            \omega_{C}&=\omega_{3l+2}=\omega_0+\Delta\cos(\phi_0+4\pi/3)
      \ .
\end{align}
This motivates us to introduce the label $l\in\{0,\ldots,L/3-1\}$ for each unit cell or trimer. Within each trimer one has three species of bosons $b_{A,l}=b_{3l}$, $b_{B,l}=b_{3l+1}$ and $b_{C,l}=b_{3l+2}$ with a similar convention for the mean fields $\alpha_{A,l}$, $\alpha_{B,l}$ 
and $\alpha_{C,l}$.
In the semi-classical limit $|\alpha_{m}|\gg1$, one can consider the effect of the quantum fluctuations at a Gaussian level, which enables us to make the decomposition 

\begin{equation}
      \label{eq:HGaussian}
	\hat{H}_{\bm{\alpha}} = \frac{L}{3}\mathcal{H}_{\text{Class}}(\bm{\alpha})+
		 \hat{H}_{\text{Lin}}(\bm{b},\bm{\alpha})+
		 \hat{H}_{\text{Quad}}(\bm{b},\bm{\alpha})
	\ ,
\end{equation}
where $\bm{b}=(b_0,\ldots,b_{L-1})$, and $\bm{\alpha}=(\alpha_0,\alpha_2,\ldots,\alpha_{L-1})$. The terms $\hat{H}_{\text{Lin}}(\bm{b},\bm{\alpha})$ and $\hat{H}_{\text{Quad}}(\bm{b},\bm{\alpha})$ are linear and quadratic in the bosonic operators, respectively. In addition, if we assume that the mean field do not depend on the position $l$ of the unit cell, we obtain the Hamilton function
\begin{align}
      \label{eq:ClassHam}
            \mathcal{H}_{\text{Class}}(\bm{\alpha}) = &\omega_A |\alpha_{A}|^2+\omega_B |\alpha_{B}|^2+\omega_C |\alpha_{C}|^2 \nonumber \\ &
            -J \left(\alpha^{*}_A\alpha_{B}+\alpha^{*}_B\alpha_{C}+\alpha^{*}_C\alpha_{A}+H.c.\right)     
            \nonumber \\ &
            +\frac{U}{2}\left[|\alpha_{A}|^2(|\alpha_{A}|^2-1)+|\alpha_{B}|^2(|\alpha_{B}|^2-1)+|\alpha_{C}|^2(|\alpha_{C}|^2-1)\right]
       \ .
\end{align}
The quantum fluctuations are governed by the quadratic Hamiltonian
\begin{align}
      \label{eq:QuadHam}
            \hat{H}_{\text{Quad}}(\bm{a},\bm{\alpha}) = &\sum^L_{m=0}\left(\omega_m-\frac{U}{2}\right)b^{\dagger}_m b_m -J \sum^{L-1}_{m=1} \left(b^{\dagger}_mb_{m+1}+h.c.\right)\nonumber \\ &+\frac{U}{2}\sum^{L}_{m=0}\left(|\alpha_{m}|^2b^{\dagger}_m b_m+\alpha_{m}^2(b^{\dagger}_m)^2+(\alpha^{*}_{m})^2b^{2}_m\right)
       \ ,
\end{align}
where $\alpha_{3l}=\alpha_{A}$, $\alpha_{3l+1}=\alpha_{B}$ 
and $\alpha_{3l+2}=\alpha_{C}$.
One can interpret the Hamiltonian $\hat{H}_{\bm{\alpha}}$ as the Hamiltonian in neighborhood of a stationary points of the energy landscape Eq.~\eqref{eq:ClassHam}. To obtain the stationary points, we require vanishing linear bosonic terms in Eq.~\eqref{eq:HGaussian}, i.e., $\hat{H}_{\text{Lin}}(\bm{a},\bm{\alpha})=0$. This conditions is satisfied as long as the mean fields $\alpha_{A},\alpha_{B}$ and $\alpha_{C}$ are a solution of the semi-classical equations of motion. The simplest solution to these equations is $\alpha_{A}=\alpha_{B}=\alpha_{C}=0$.
In this case, the Hamiltonian in Eq.~\eqref{eq:QuadHam} takes a simple form
\begin{equation}
      \label{eq:QuadHamZeroMeanField}
            \hat{H}_{\text{Quad}}(\bm{a},\bm{\alpha}) = \sum^{L/3-1}_{l=0}(\bm{\Psi}_l^{\dagger})^{\text{T}}\bm{\mathcal{M}}\bm{\Psi}_l+\sum^{L/3-2}_{l=0}J\left[(\bm{\Psi}_l^{\dagger})^{\text{T}}\bm{\mathcal{N}}\bm{\Psi}_{l+1}+H.c\right]
      \ ,
\end{equation}
where $(\bm{\Psi}_l^{\dagger})^{\text{T}}=(b^{\dagger}_{A,l},b^{\dagger}_{B,l},b^{\dagger}_{C,l})$. Correspondingly, the matrices are 
\begin{align}
       \label{eq:LocalFreq}
       \boldsymbol{\mathcal{M}} &=\left(%
\begin{array}{ccc}
\omega_{A}-\frac{U}{2}  &  0 &  0  \\
 0 & \omega_{B}-\frac{U}{2} &  0 \\
 0 &    0 &  \omega_{C}-\frac{U}{2}
\end{array}
\right)
\ ,
\end{align}
and $\bm{\mathcal{N}}=\text{diag}(1,1,1)$.

We introduce here a discrete Fourier transformation $\boldsymbol{\Psi}_{l}=\sqrt{\frac{3}{L}}\sum_{k}\boldsymbol{\Phi}_{k} e^{\mathrm{i} k l}$, where
$(\boldsymbol{\Phi}^{\dagger}_{k})^{T}=(b^{\dagger}_{A,k},b^{\dagger}_{B,k},b^{\dagger}_{C,k})$, and $b_{\mu,l}=\sqrt{\frac{3}{L}}\sum_{k}b_{\mu,k} e^{\mathrm{i} k l}$ with $\mu\in\{A,B,C\}$.
Now we can write Hamiltonian Eq.~\eqref{eq:QuadHamZeroMeanField} as
$\hat{H}_{\text{Quad}}(\bm{a},\bm{\alpha})=\sum_{k}(\boldsymbol{\Phi}^{\dagger}_{k})^{T}\boldsymbol{H}_{k}\boldsymbol{\Phi}_{k}$ with the Bogoliubov de Gennes Hamiltonian $\boldsymbol{H}_{k}=\boldsymbol{\mathcal{M}}+2J \boldsymbol{\mathcal{N}}\cos k$. Finally, by considering $U=-J$, we arrive to the dispersion energies $E_{\mu,k}=\omega_\mu+J/2+2J\cos k$. From this, one can see that when $\omega_A=\omega_B$ or $\omega_A=\omega_C$, we obtain gapless excitations and therefore, a quantum phase transition.

\subsection{Circuit QED implementation\label{AppendixC}}

In this section, we review how an array of capacitively-coupled transmon qubits \cite{transmon} can be mapped to a nonlinear coupled resonator array, as described by the Hamiltonian in Eq.~[1]. The derivation provided here can be generalized to a more complex coupler such as a transmission line \cite{cqed_hartmann, transmissionline1,transmissionline2} and an inductive tunable coupler \cite{gmoncoupler}.

\begin{figure}
\includegraphics[width=10cm,height=3.5cm]{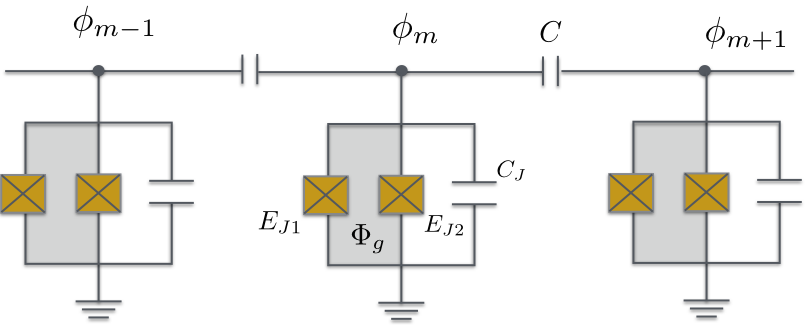}
\caption{Circuit diagram implementing the nonlinear coupled resonator array discussed in the main text.}
\label{fig:figa2}
\end{figure}

Our circuit diagram is shown in Fig.\ref{fig:figa2}. The flux variable is defined as $\phi_m=-\int V_m dt$, where $V_m$ is a voltage at the corresponding position. As will be shown below, this quantity can be quantized to the form $\phi_m=\alpha (a_m+a^\dagger_m)$, where $a_m, a_m^\dagger$ are bosonic operators of an ''artificial'' photon at site $m$ and $\alpha$ is some constant that depends on the circuit's elements. As shown in \cite{transmon}, two parallel-connected Josephson junction with a flux bias $\Phi_g$ can be thought of as an effective single Josephson junctions $E_J$ where 
\begin{equation}
E_{J}=(E_{J1}+E_{J2})\cos\left(\frac{\Phi_g}{2\Phi_0}\right)\sqrt{1+d^2\tan \left(\frac{\Phi_g}{2\Phi_0}\right)},
\end{equation}
with $\Phi_0=\hbar/2e$ and $d=(E_{J2}-E_{J1})/(E_{J2}+E_{J1})$. The resonator's frequency $\omega_m$ is related to $E_J$, hence it can be tuned on the fly, by changing the flux bias $\Phi_g$.

Following the standard circuit quantization procedure \cite{Devoret_lecture}, we first write down the circuit's Lagrangian as 
\begin{equation}
\mathcal{L}=\sum_{m=0}^{L-1}\left(\frac{1}{2}C_J\dot{\phi}_m^2+E_J\cos\left(\frac{\phi_m}{\phi_0}\right)\right)+\sum_{m=0}^{L-2}\frac{1}{2}C(\dot{\phi}_m-\dot{\phi}_{m+1})^2,
\end{equation}
Assuming $C/(C_J+2C)\ll 1$, the Hamiltonian can be obtained using the Legendre transformation \cite{Nunnenkamp},
\begin{equation}
H=\sum_{m=0}^{L-1}\left(\frac{\dot{\phi}_m^2}{2\tilde{C}}+\frac{\phi_m^2}{2\tilde{L}}+\sum_{n=2}^{\infty}\frac{(-1)^nE_J}{(2n)!\Phi_0^{2n}}\phi_m^{2n}\right)+\sum_{m=0}^{L-2}\frac{C}{\tilde{C}^2}q_mq_{m+1},
\label{eqApp:H}
\end{equation}
where $q_m=\sqrt{2C+C_J}\partial \mathcal{L}/\partial \dot{\phi_m}$ is a conjugate momentum of $\phi_m$, $\tilde{C}= C_J+2C$ is an effective capacitance and $\tilde{L}=\Phi_0^2/E_J$ is an effective inductance . We then quantized $\phi_m$ and $q_m$ by defining ladder operators $a_m$, $a^\dagger_m$ according to $\phi_m  = (\tilde{L}/4\tilde{C})^{1/4}(a_m+a^\dagger_m)$ and $q_m = i(\tilde{C}/4\tilde{L})^{1/4}(-a_m+a^\dagger_m)$. The first two terms in Eq.\ref{eqApp:H} become $\sum_m\omega a^\dagger_ma_m$, where $\omega=1/\sqrt{\tilde{L}\tilde{C}}$ is a resonator frequency. In addition, the capacitor $C$ leads to the hopping term with $J=-\frac{\omega C}{2\tilde{C}}$. A rotating-wave approximation is assumed, so we ignore the term $(a^\dagger_ma^\dagger_{m+1}+h.c.)$.

The Josephson junction $E_J$ introduces an anharmonicity to the resonator's frequency. Due to this anharmonicity, a vacuum state $|0\rangle$ and a one-photon Fock state $|1\rangle$ of the resonator can be used as a qubit. A transmon qubit corresponds to the regime with a large $E_{\tilde{L}}/E_{\tilde{C}}>1$ where $E_{\tilde{C}}=e^2/2\tilde{C}$ and $E_{\tilde{L}}=\phi_0^2/\tilde{L}$ , such that the terms higher than the forth order can be neglected\cite{transmon}. Hence, a transmon qubit can be thought of as a resonator with an attractive Kerr nonlinearity $U<0$. Taking into account the normal ordering \cite{cqed_hartmann}, we get $U=-E_Je^{-\lambda^2}\lambda^4/4$, where $\lambda =(2E_{\tilde{C}}/E_{\tilde{L}})^{1/4}$. This normal ordering also introduces a small normalisation factor $\delta \omega$to the resonator frequency, with $\delta \omega = \lambda^2E_Je^{-\lambda^2}$.

Note that all Hamiltonian parameters ($\omega_m$, $J$ and $U$) depend on $E_J$. Hence, in general, tuning $\omega$ also effects other parameters as well. In the main text, we tune the resonator frequency within the range $\left[\omega_0-\Delta,\omega_0+\Delta \right]$, where $\omega_0\sim 5$ GHz, $\Delta = 400$ MHz and $J=-U=40$ MHz. Hence $\omega_m$ only changes by $\sim 8\%$. Therefore, subsequent changes in $J$ and $U$ are small compared to $\Delta$ and do not alter our discussion in the main text.

\end{widetext}

\bibliographystyle{apsrev4-1}
\bibliography{Xbib,ref}

\end{document}